\documentclass[aps,prl,twocolumn,showpacs,superscriptaddress]{revtex4-1}
\usepackage{graphicx}

\begin{document}

\title{Electron-doping evolution of the low-energy spin excitations in the
iron arsenide BaFe$_{2-x}$Ni$_{x}$As$_{2}$ superconductors}
\author{Miaoyin Wang}
\affiliation{Department of Physics and Astronomy, The University of Tennessee, Knoxville,
Tennessee 37996-1200, USA }
\author{Huiqian Luo}
\affiliation{Beijing National Laboratory for Condensed Matter Physics and Institute of
Physics, Chinese Academy of Sciences, P. O. Box 603, Beijing 100190, China }
\author{Jun Zhao}
\affiliation{Department of Physics and Astronomy, The University of Tennessee, Knoxville,
Tennessee 37996-1200, USA }
\author{Chenglin Zhang}
\affiliation{Department of Physics and Astronomy, The University of Tennessee, Knoxville,
Tennessee 37996-1200, USA }
\author{Meng Wang}
\affiliation{Beijing National Laboratory for Condensed Matter Physics and Institute of
Physics, Chinese Academy of Sciences, P. O. Box 603, Beijing 100190, China }
\affiliation{Department of Physics and Astronomy, The University of Tennessee, Knoxville,
Tennessee 37996-1200, USA }
\author{Karol Marty}
\affiliation{Neutron Scattering Science Division, Oak Ridge National Laboratory, Oak
Ridge, Tennessee 37831-6393, USA}
\author{Songxue Chi}
\affiliation{NIST Center for Neutron Research, National Institute of Standards and
Technology, Gaithersburg, MD 20899, USA }
\author{Jeffrey W. Lynn}
\affiliation{NIST Center for Neutron Research, National Institute of Standards and
Technology, Gaithersburg, MD 20899, USA }
\author{Astrid Schneidewind}
\affiliation{Gemeinsame Forschergruppe HZB - TU Dresden, Helmholtz-Zentrum Berlin f$\ddot{%
u}$r Materialien und Energie, D-14109 Berlin, Germany }
\affiliation{Forschungsneutronenquelle Heinz Maier-Leibnitz (FRM-II), TU M$\ddot{u}$%
nchen, D-85747 Garching, Germany}
\author{Shiliang Li}
\email{slli@aphy.iphy.ac.cn}
\affiliation{Beijing National Laboratory for Condensed Matter Physics and Institute of
Physics, Chinese Academy of Sciences, P. O. Box 603, Beijing 100190, China }
\author{Pengcheng Dai}
\email{daip@ornl.gov}
\affiliation{Department of Physics and Astronomy, The University of Tennessee, Knoxville,
Tennessee 37996-1200, USA }
\affiliation{Beijing National Laboratory for Condensed Matter Physics and Institute of
Physics, Chinese Academy of Sciences, P. O. Box 603, Beijing 100190, China }
\affiliation{Neutron Scattering Science Division, Oak Ridge National Laboratory, Oak
Ridge, Tennessee 37831-6393, USA}

\begin{abstract}
We use elastic and inelastic neutron scattering to systematically
investigate the evolution of the low-energy spin excitations of the iron
arsenide superconductor BaFe$_{2-x}$Ni$_{x}$As$_{2}$ as a function of nickel
doping $x$. In the undoped state, BaFe$_{2}$As$_{2}$ exhibits a
tetragonal-to-orthorhombic structural phase transition and simultaneously
develops a collinear antiferromagnetic (AF) order below $T_{N}=143$ K. Upon
electron-doping of $x=0.075$ to induce bulk superconductivity with $T_{c}=12.3$ K, the AF ordering temperature reduces to $T_{N}\approx 58$ K.
We show that the appearance of bulk superconductivity in BaFe$_{1.925}$Ni$_{0.075}$As$_{2}$ coincides with a dispersive neutron spin resonance in the
spin excitation spectra, and a reduction in the static ordered moment. For
optimally doped BaFe$_{1.9}$Ni$_{0.1}$As$_{2}$ ($T_{c}=20$ K) and overdoped
BaFe$_{1.85}$Ni$_{0.15}$As$_{2}$ ($T_{c}=15$ K) superconductors, the static
AF long-range order is completely suppressed and the spin excitation spectra
are dominated by a resonance and spin-gap at lower energies. We determine
the electron-doping dependence of the neutron spin resonance and spin gap
energies, and demonstrate that the three-dimensional nature of the resonance
survives into the overdoped regime. If spin excitations are important for
superconductivity, these results would suggest that the three-dimensional
character of the electronic superconducting gaps are prevalent throughout
the phase diagram, and may be critical for superconductivity in these
materials.
\end{abstract}

\pacs{74.25.Ha, 74.70.-b, 78.70.Nx}

\maketitle
\section{I. Introduction}
An experimental determination of the doping evolution of the spin
excitations in iron arsenide superconductors \cite{kamihara,rotter,ljli,budko} is important for a comprehensive understanding
of the role of magnetism in the superconductivity of these materials. Like
high-transition temperature (high-$T_{c}$) copper oxides, the parent
compounds of iron arsenide superconductors exhibit static antiferromagnetic
(AF) long-range order with a collinear spin structure \cite{cruz,jzhao1,huang,zhao2}. Although there is currently no consensus on a
microscopic mechanism for superconductivity, spin excitations have been
postulated by several theories to play a crucial role in the electron
pairing and superconductivity of these materials \cite{mazin,chubkov,fwang,cvetkovic,moreo}. In one class of unconventional
microscopic theories for superconductivity, electron pairing in iron
arsenide superconductors is mediated by quasiparticle excitations between
sign reversed hole pockets around the $\Gamma $ point and the electron
Fermi pockets around the $M$ point as shown in the inset of Fig. 1(a) \cite{mazin1}. If
this is indeed the case, spin excitations in the superconducting state
should have a collective mode called the neutron spin resonance, whose
energy is at (or slightly less than) the addition of the hole and electron
superconducting gap energies ($E =\left\vert \Delta (k+Q)+\Delta
(k)\right\vert $, where $Q$ is the AF ordering wavevector connecting the
hole and electron Fermi pockets at the $\Gamma $ and $M$ points,
respectively) \cite{maier1,maier2,korshunov,seo09}. Although recent
inelastic neutron scattering experiments have found the neutron spin
resonance for different iron-based superconductors consistent with this
picture \cite{christianson,lumsden,chi,slli,parshall,pratt,christianson09,inosov}, a
surprising result has been that the mode in the optimally doped BaFe$_{1.9}$Ni$_{0.1}$As$_{2}$ ($T_{c}=20$ K) has three-dimensional character with clear
dispersion along the $c$-axis \cite{chi,slli}, quite different from the
two-dimensional nature of the resonance in copper oxide superconductors \cite{mignod,dai,fong,stock,wilson,jzhao07}. 
If spin excitations are important
for superconductivity in iron-arsenides, it would be interesting to
systematically investigate the doping evolution of the resonance in BaFe$_{2-x}$Ni$_{x}$As$_{2}$, and determine if the three-dimensional nature of
the mode is a general phenomenon or specific only to the optimally doped
materials. Furthermore, since spin waves in the AF ordered parent compounds
of (Ba,Sr,Ca)Fe$_{2}$As$_{2}$ have rather large anisotropy spin gaps at the
AF zone center \cite{jzhao2,ewings,rob,matan,diallo,jzhao3} and the
optimally doped superconducting samples are generally gapless \cite{lumsden,chi,slli,inosov}, it would be important to see how spin waves in
the parent compounds evolve as electrons are doped into the FeAs planes.

\begin{figure}[t]
\includegraphics[scale=.4]{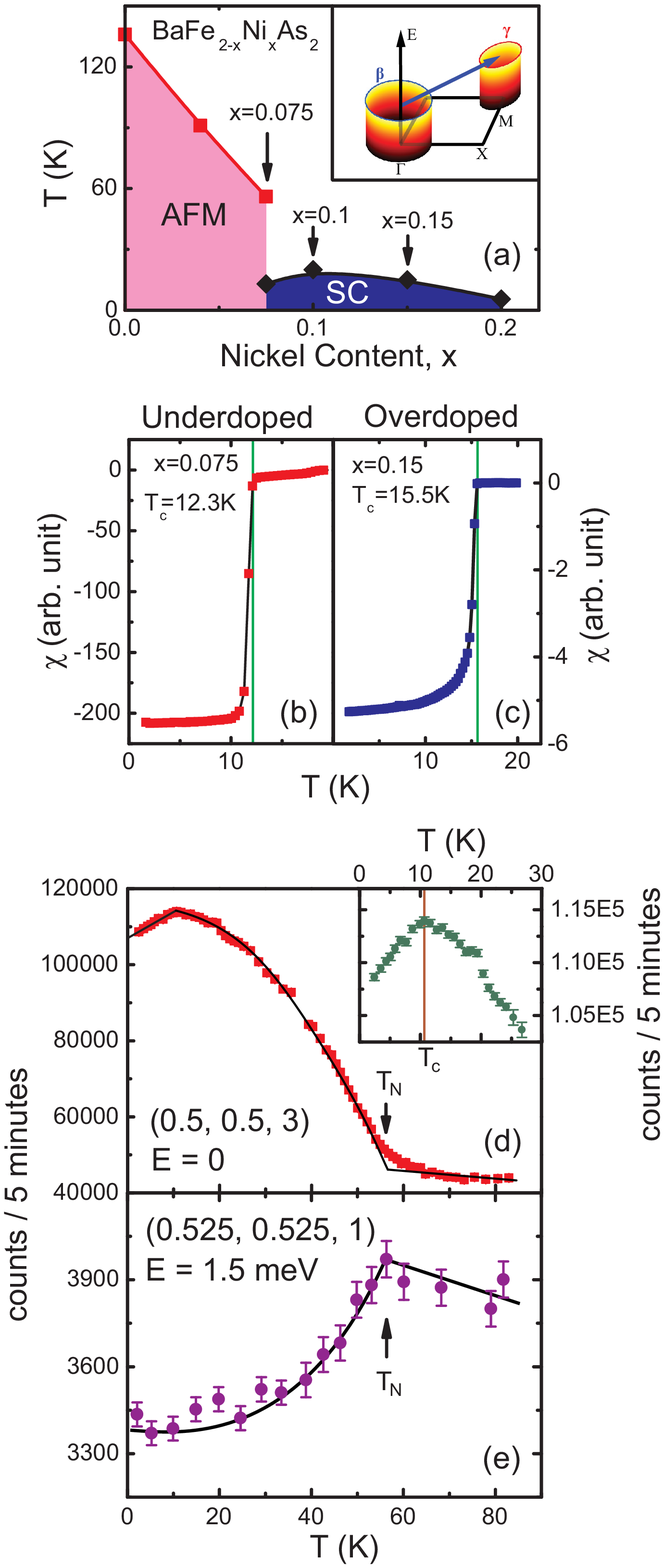}
\caption{ (color online) (a) Electronic phase diagram of BaFe$_{2-x}$Ni$_{x}$
As$_{2}$ as determined from our previous \protect\cite{chi,slli,harriger}
and current neutron scattering work. The inset shows schematic illustration of
quasiparticle excitations from the hole pocket at the $\Gamma $ point to the
electron pocket at the $M$ point as predicted by various theories \cite{maier1,maier2,korshunov,seo09}. 
(b,c) Temperature dependence of the Meissner and shielding signals on small
crystals of BaFe$_{1.925}$Ni$_{0.075}$As$_{2}$ and BaFe$_{1.85}$Ni$_{0.15}$As
$_{2}$.
(d) Temperature dependence of the 
$Q=(0.5,0.5,3)$ magnetic Bragg peak in BaFe$_{1.925}$Ni$_{0.075}$As$_{2}$,
showing a clear anomaly at $T_{N}$ and $T_{c}$. The inset shows an expanded
view of the $Q=(0.5,0.5,3)$ magnetic Bragg peak near $T_{c}$.  (e) Temperature 
dependence of the quasi-elastic scattering at $E =1.5$ meV and $Q=(0.525,0.525,1)$ shows a clear anomaly at $T_N=58$ K.}
\end{figure}

\begin{figure}[t]
\includegraphics[scale=.4]{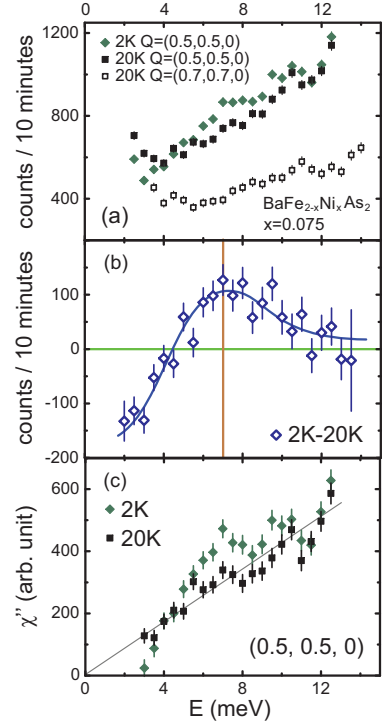}
\caption{
(color online) (a) Energy scans at $Q = (0.5,0.5,0)$ (signal) and $Q=(0.7,0.7,0)$ (background) positions 
above and below $T_{c}$ for BaFe$_{1.925}$Ni$_{0.075}$As$_2$.
(b) The temperature difference plot (low temperature minus high temperature) 
in $S(Q,\omega)$ shows a clear neutron spin resonance at $E=7$ meV below $T_c$.
The solid line is a guide to the eye. (c)
Estimation of the temperature dependence of the $\chi^{\prime\prime}(Q,\omega)$ above and below $T_c$,
using background determined both from constant-energy scans and from constant-$Q$ scans at background position.
}
\end{figure}

In this article, we report our inelastic neutron scattering studies of the
low-energy spin excitations in electron-doped BaFe$_{2-x}$Ni$_{x}$As$_{2}$
with $x=0.075,0.15$ [Fig. 1(a)], and compare and contrast them with the spin
excitations observed at optimal doping and in the lightly/undoped regime.
Before Ni-doping, BaFe$_{2}$As$_{2}$ exhibits simultaneous structural and
magnetic phase transitions below $T_{s}=T_{N}=143$ K, changing the crystal
lattice symmetry from the high-temperature tetragonal to the low-temperature
orthorhombic phase \cite{huang}. Upon doping electrons via either Co or Ni
substitution for Fe, the structural and magnetic phase transitions are
separated \cite{chu,lester}. For $x=0.04$ Ni-doping, the structural and
magnetic phase transition temperatures of BaFe$_{1.96}$Ni$_{0.04}$As$_{2}$
become near 97 K and 91 K, respectively \cite{harriger}. In addition, three
dimensional spin waves in BaFe$_{2}$As$_{2}$ change into
quasi-two-dimensional spin excitations for BaFe$_{1.96}$Ni$_{0.04}$As$_{2}$
with no evidence for the neutron spin resonance \cite{harriger} or bulk
superconductivity \cite{budko}. By increasing the Ni-doping $x$ to 0.075 to
form BaFe$_{1.925}$Ni$_{0.075}$As$_{2}$ [Fig. 1(a)], bulk superconductivity
appears at $T_{c}=12.3$ K [Fig. 1(b)] and the N$\mathrm{\acute{e}e}$l
temperature of the material is now reduced to $T_{N}\approx 58$ K [Figs.
1(d) and 1(e)]. Our inelastic neutron scattering experiments show the presence of a
three-dimensional neutron spin resonance with distinct energies at the AF
wavevectors $Q=(0.5,0.5,0)$ and $(0.5,0.5,1)$, quite similar to that of the
optimally doped BaFe$_{1.9}$Ni$_{0.1}$As$_{2}$ \cite{chi,slli}. The
intensity gain of the mode below $T_{c}$ is compensated by opening a pseudo
spin gap at lower energies and reduction in the static AF order [see inset in Fig. 1(d)].

To study the doping evolution of the resonance, we also carried out
inelastic neutron scattering experiments on overdoped BaFe$_{1.85}$Ni$_{0.15}
$As$_{2}$ [$T_{c}=15.5$ K, Fig. 1(c)] and found that the energy of the mode
is approximately proportional to $T_{c}$. Our elastic neutron scattering
measurements indicate that the static antiferromagnetism has been completely
suppressed, while the neutron spin resonance in the superconducting state
exhibits similar dispersion along the $c$-axis as the underdoped and
optimally doped materials. This suggests that the three-dimensional nature
of the resonance energy and its associated superconducting gap energy $\Delta $ are prevalent throughout the superconducting electronic phase
diagram. These results can also provide information needed for calculating
the electron-doping dependence of the AF coupling between the layers, and
estimating the doping dependence of the superconducting gap energy.

\begin{figure}[t]
\includegraphics[scale=.4]{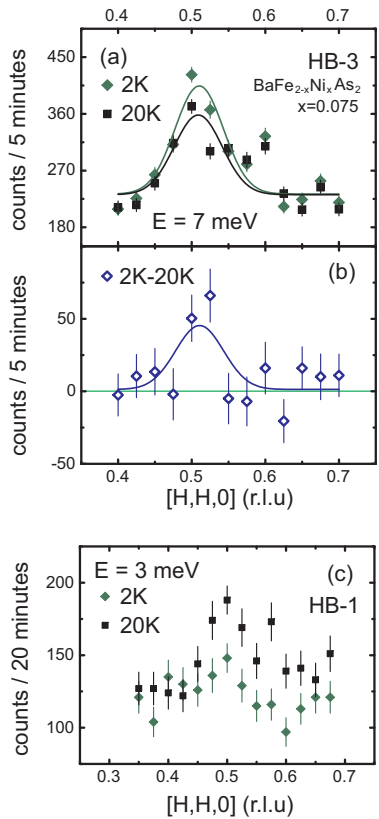}
\caption{(color online)
(a) Constant-energy scans near the resonance energy ($E=7$ meV)
along the $(H,H,0)$ direction across $T_c$ for BaFe$_{1.925}$Ni$_{0.075}$As$_2$. 
The scattering clearly increases at the
$Q=(0.5,0.5,0)$ below $T_c$.  The small peak at $Q=(0.6,0.6,0)$ is spurious.
(b) Temperature difference plot confirms the intensity gain at $Q=(0.5,0.5,0)$ below $T_c$.
(c) Wave-vector scans at $E=3$ meV.  Here the scattering at $Q=(0.5,0.5,0)$ 
decreases below $T_c$ but there is no spin gap at $E=3$ meV.
 }
\end{figure}

\begin{figure}[t]
\includegraphics[scale=.4]{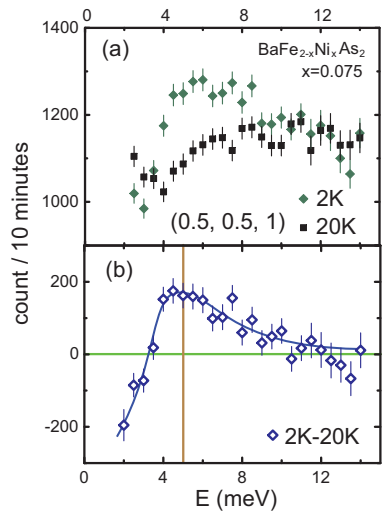}
\caption{(color online)
(a) Constant-$Q$ scans at $Q = (0.5,0.5,1)$  
above and below $T_{c}$ for BaFe$_{1.925}$Ni$_{0.075}$As$_2$.  The scattering shows clear asymmetric 
enhancement around $E=5$ meV below $T_c$. (b) Temperature difference plot reveals a neutron spin
resonance at $E=5$ meV, clearly below the energy of the mode at $Q = (0.5,0.5,0)$ as shown 
in Fig. 2.
 }
\end{figure}

\section{II. Experimental Details}

In two recent inelastic neutron scattering experiments on underdoped BaFe$_{1.906}$Co$_{0.094}$As$_{2}$ ($T_{c}=15$ K) \cite{pratt} and BaFe$_{1.92}$Co$_{0.08}$As$_{2}$ ($T_{c}=11$ K) \cite{christianson09}, static AF order was
found to coexist with superconductivity and cooling below $T_{c}$'s in these
samples induced a weak neutron spin resonance in the magnetic excitation
spectra at the expense of elastic magnetic scattering \cite{pratt,christianson09}. For BaFe$_{2-x}$Ni$_{x}$As$_{2}$, bulk
superconductivity appears only when $x\geq 0.05$ \cite{budko}. To compare
the electronic phase diagram of BaFe$_{2-x}$Ni$_{x}$As$_{2}$ with Co-doped
materials and see the effect of superconductivity on the spin excitations,
we chose to study underdoped BaFe$_{1.925}$Ni$_{0.075}$As$_{2}$ (where Ni
concentration is nominal) and overdoped BaFe$_{1.85}$Ni$_{0.15}$As$_{2}$
superconductors. The temperature dependence of the susceptibility in Figures
1(b) and 1(c) show $T_{c}$'s of 12.3 K and 15.5 K for BaFe$_{1.925}$Ni$_{0.075}$As$_{2}$ and BaFe$_{1.85}$Ni$_{0.15}$As$_{2}$, respectively,
consistent with the overall electronic phase diagram from heat capacity
measurements \cite{budko}.

We grew single crystals of BaFe$_{2-x}$Ni$_{x}$As$_{2}$ with $x=0.075,0.15$
using the self-flux method \cite{ljli}. Our neutron scattering experiments
were carried out on the HB-3, HB-1 thermal triple-axis spectrometers at the
high-flux-isotope reactor (HFIR), Oak Ridge National Laboratory \cite{jzhao2}; the BT-7 thermal triple-axis spectrometer at the NIST Center for Neutron
Research \cite{slli}; and the PANDA cold triple-axis spectrometer at the
Forschungsneutronenquelle Heinz Maier-Leibnitz (FRM-II), TU M$\mathrm{\ddot{u}}$nchen \cite{chi}. 
We defined the wave vector $Q$ at ($q_{x}$, $q_{y}$, $q_{z}$) as $(H,K,L)=(q_{x}a/2\pi ,q_{y}b/2\pi ,q_{z}c/2\pi )$ reciprocal
lattice units (rlu) using the tetragonal nuclear unit cell, where $a=3.89$ 
\AA , $b=3.89$ \AA , and $c=12.77$ \AA . We co-aligned about 6 grams for
each of the $x=0.075,0.15$ samples of BaFe$_{2-x}$Ni$_{x}$As$_{2}$ in the $[H,H,L]$ horizontal scattering plane, and put our samples inside either a
closed cycle refrigerator or a liquid He cryostat.

For thermal triple-axis measurements on HB-1, HB-3, and BT-7, we used
pyrolytic graphite (PG) as monochromator and analyzer with typical
collimations of open-40$^{\prime \prime }$-S-40$^{\prime \prime }$-120$^{\prime \prime }$. The final neutron energy was chosen to be either $E_{f}=13.5$
meV or $E_{f}=14.7$ meV with a PG filter before the analyzer. For cold
triple-axis measurements on PANDA, we chose final neutron energy of $E_{f}=5.0$ meV with a cooled Be filter in front of the analyzer. We used both
horizontal and vertical focusing PG monochromator and analyzer with no
collimators. We also used a $E_{f}=13.5$ meV setup with a PG filter in one
of the PANDA measurements.

\begin{figure}[t]
\includegraphics[scale=.4]{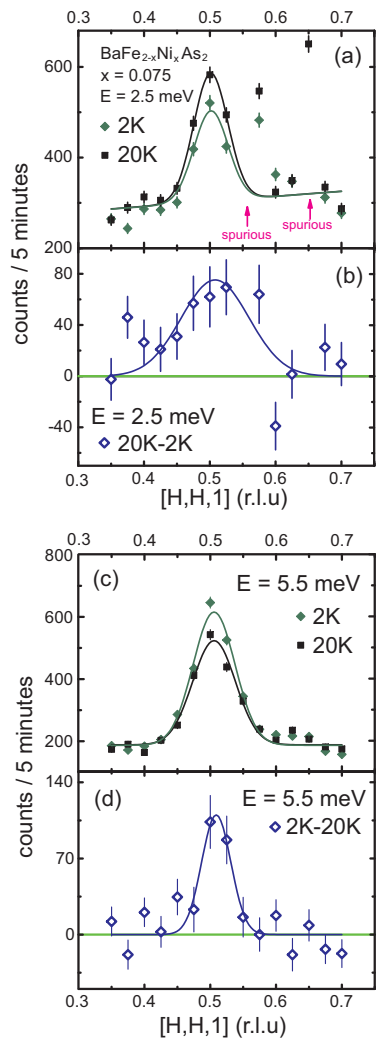}
\caption{(color online)
(a) $Q$-scans along the $(H,H,1)$ direction at $E=2.5$ meV above and below $T_c$
for BaFe$_{1.925}$Ni$_{0.075}$As$_2$.
(b) Temperature difference plot shows that the scattering at $Q=(0.5,0.5,1)$ 
and $E=2.5$ meV decreases below $T_c$.
(c) Identical $Q$-scans across $T_c$ at the resonance energy of $E=5.5$ meV.
The scattering enhances below $T_c$. (d) Temperature difference plot between 2 K and 20 K, 
showing clear field-induced scattering below $T_c$ at $Q=(0.5,0.5,1)$.
 }
\end{figure}

\section{III. Results and Discussions}

We first describe our elastic and quasielastic neutron scattering results on
the underdoped BaFe$_{1.925}$Ni$_{0.075}$As$_{2}$. Consistent with earlier
results on underdoped BaFe$_{1.906}$Co$_{0.094}$As$_{2}$ \cite{pratt} and
BaFe$_{1.92}$Co$_{0.08}$As$_{2}$ \cite{christianson09}, the AF structure of
the BaFe$_{1.925}$Ni$_{0.075}$As$_{2}$ sample reported here is identical to
the undoped parent compound but with a N$\mathrm{\acute{e}e}$l $T_{N}\approx
58$ K [Fig. 1(c)]. The temperature dependence of the quasielastic scattering
at $Q=(0.525,0.525,1)$ and $E=1.5$ meV shows a clear kink below $\sim 58$ K,
thus confirming the N$\mathrm{\acute{e}e}$l temperature of the system. The
inset in Figure 1(d) shows the expanded temperature dependence of the AF
Bragg peak intensity at $Q=(0.5,0.5,3)$. The scattering decreases with
decreasing temperature at the onset of $T_{c}$, suggesting that the static
moment competes with superconductivity similar to the Co-doped materials 
\cite{pratt,christianson09}.

\begin{figure}[t]
\includegraphics[scale=.5]{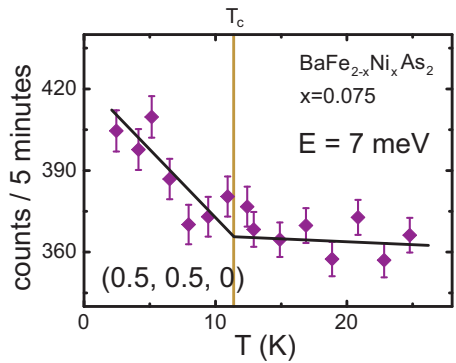}
\caption{(color online)
(a) Temperature dependence of the $E=7$ meV scattering at 
$Q=(0.5,0.5,0)$ for BaFe$_{1.925}$Ni$_{0.075}$As$_2$.  The scattering increases in intensity below $T_c$ of
12.3 K.
 }
\end{figure}

To see if there is a neutron spin resonance mode in underdoped BaFe$_{1.925}$Ni$_{0.075}$As$_{2}$ and to compare its $c$-axis dispersion with optimally
doped BaFe$_{1.9}$Ni$_{0.1}$As$_{2}$ \cite{chi}, we carried out constant-$Q$
scans at $Q=(0.5,0.5,0)$ and $(0.5,0.5,1)$ above and below the
superconducting transition temperature $T_{c}$. Figure 2(a) shows the raw
data collected on the HB-3 triple-axis spectrometer at the signal $Q=(0.5,0.5,0)$ and background $Q=(0.7,0.7,0)$ positions. There is clear
intensity gain at $Q=(0.5,0.5,0)$ near $E=7$ meV below $T_{c}$ at the
expense of spectral weight loss below $\sim $4 meV. The temperature
difference spectrum between 2 K and 20 K in Fig. 2(b) confirms the presence
of the mode at $E =7$ meV below $T_{c}$ and a reduction in
spectral weight below 4 meV. The open squares in Fig. 2(a) show the energy
dependence of the background scattering at $Q=(0.7,0.7,0)$.

\begin{figure}[t]
\includegraphics[scale=.35]{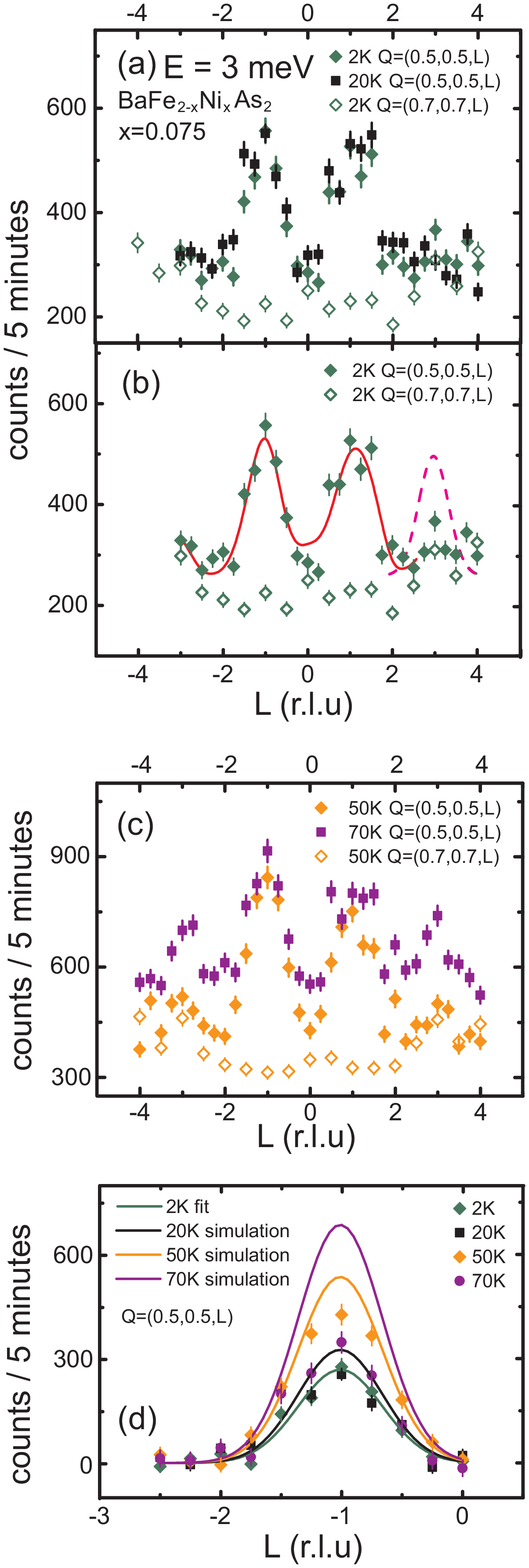}
\caption{(color online)
Constant-energy scans along the $(0.5,0.5,L)$ (signal) and $(0.7,0.7,L)$ (background) directions
at $E=3$ meV and various temperatures. (a) The scattering at 2 K and 20 K shows well-defined
peaks centered at $L=\pm 1$. Fourier transform of these peaks give $c$-axis spin-spin correlation length 
of $\sim14$ \AA. (b) Signal and background Scattering at 2 K.  The solid line is Gaussian fits to the
data.  The dashed line shows the expected magnetic scattering at $Q=(0.5,0.5,3)$ assuming Fe$^{2+}$ 
form factor.  The absence of clear peaks at $L=\pm 3$ suggests that magnetic scattering in 
BaFe$_{1.925}$Ni$_{0.075}$As$_2$ damps out much faster than expected.
(c) Similar scans at 50 K and 70 K. (d) The solid lines in the Figure
 show the effect of Bose population factor as a function of increasing temperature if one normalizes the
 magnetic scattering above background at 2 K.  The magnetic intensity changes in the system clearly does not
 obey the Bose statistics, indicating that the scattering is not spin waves. 
}
\end{figure}

Figures 3 summarizes constant-energy scans at $E =7$ meV and 3
meV along the $[H,H,0]$ direction. At the resonance energy, the scattering
shows a well-defined peak centered at $Q=(0.5,0.5,0)$ that increases in
intensity below $T_{c}$ [Fig. 3(a)]. Figure 3(b) shows the temperature
difference plot which confirms that the intensity gain below $T_{c}$ in Fig. 2(a) occurs at $Q=(0.5,0.5,0)$. 
Similarly, constant-energy scans at $E =3$ meV above and below $T_{c}$ in Fig. 3(c) reveal clear normal
state magnetic scattering that is not completely suppressed (gapped) below $T_{c}$, 
at least with the energy resolution afforded with these thermal
triple-axis measurements. Figure 2(c) shows our estimation of the energy
dependence of the dynamic susceptibility $\chi ^{\prime \prime }(Q,\omega)$
above and below $T_{c}$, obtained by subtracting the background and
correcting for the Bose population factor using $\chi ^{\prime \prime
}(Q,\omega )=[1-\exp (-\hbar \omega /k_{B}T)]S(Q,\omega )$, where $E=\hbar\omega$.  While the normal
state susceptibility appears to increase linearly with energy,
superconductivity rearranges the spectrum, creating a (pseudo) spin gap
below 4 meV and a neutron spin resonance at $E =7$ meV for
in-phase spin fluctuations along the $c$-axis ($L=0$).

To investigate the behavior for the out-of-phase spin fluctuations along the 
$c$-axis, we plot in Fig. 4(a) constant-$Q$ scans at $Q=(0.5,0.5,1)$ above
and below $T_{c}$. Figure 4(b) shows the temperature difference plot, and a
comparison of Fig. 4(b) and Fig. 2(b) immediately reveals that the neutron
spin resonance has moved from $E=7$ meV at $Q=(0.5,0.5,0)$ to $E=5$ meV at $Q=(0.5,0.5,1)$. Note in particular that for $Q=(0.5,0.5,0) $ there is essentially no change with temperature for the
scattering at $5$ meV (Fig. 2), which is where the maximum in intensity
occurs at the $Q=(0.5,0.5,1)$ position. This is compelling evidence that
the neutron spin resonance is dispersive for both underdoped and optimally
doped BaFe$_{2-x}$Ni$_{x}$As$_{2}$.

\begin{figure}[t]
\includegraphics[scale=.35]{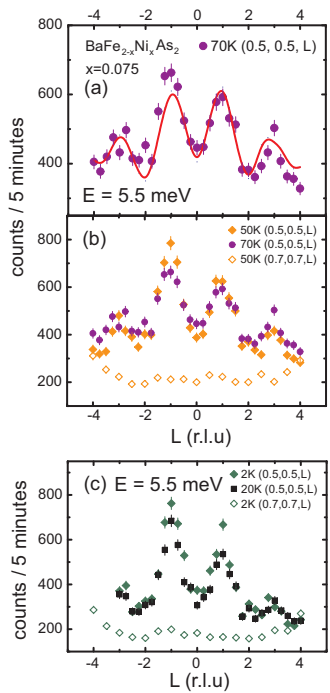}
\caption{(color online)
Constant-energy scans along the $(0.5,0.5,L)$ (signal) and $(0.7,0.7,L)$ (background) directions
at the resonance energy of $E=5.5$ meV and various temperatures.
(a) The $c$-axis scattering at 70 K.  (b) The signal and background scattering at
50 K.  The data again show clear peaks at $L=\pm 1$ but much damped peaks at 
$L=\pm 3$. (c) Signal and background scattering at 2 K and 20 K.  Superconductivity clearly
enhances magnetic scattering at $L=0$ and $L=\pm 1$.  The spin-spin correlation length
is again about 14 \AA\ and is weakly temperature dependent 
in the probed temperature range (2 K to 70 K).
 }
\end{figure}

\begin{figure}[t]
\includegraphics[scale=.35]{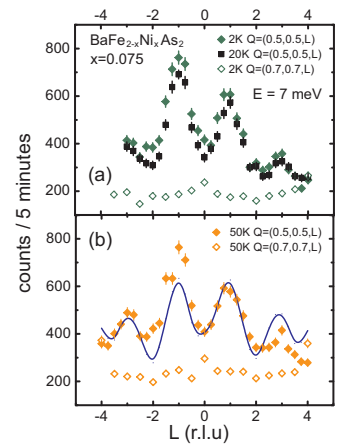}
\caption{(color online) $L$-dependence of the magnetic scattering for BaFe$_{1.925}$Ni$_{0.075}$As$_2$. 
(a) Identical scans as Fig. 8 except we now change the excitation energy to $E=7$ meV.
Comparison of the 2 K and 20 K data here indicates that the magnetic scattering
enhancement below $T_c$ at $L=0$ is larger than that at $L=1$.
(b) Signal and background scattering at 50 K.  The solid line shows the expected $L$-dependence of
the magnetic scattering
assuming simple Fe$^{2+}$ form factor, which clearly fails to describe the data.
 }
\end{figure}

Figure 5 shows constant-energy scans along the $[H,H,1]$ direction and the
temperature difference data between 2 K and 20 K for $E =2.5,5.5$
meV. Similar to the $[H,H,0]$ scans in Fig. 3, we find that
superconductivity only reduces but does not completely suppress the magnetic
scattering at $E =2.5$ meV [Figs. 5(a) and 5(b)]. Similarly, the
scattering near the resonance energy at $E =5.5$ meV shows a clear
increase below $T_{c}$. To test if the intensity gain at $E =7$
and $Q=(0.5,0.5,0)$ is responding to superconductivity, we show in Fig. 6
the temperature dependence of the scattering, which clearly increases below $T_{c}$ consistent with the temperature dependence of the neutron spin
resonance \cite{christianson,lumsden,chi,slli,pratt,christianson09,inosov}.

To determine the $c$-axis modulation of the spin excitations at different
temperatures and energies, we show in Figs. 7, 8, and 9 constant-energy
scans along the $Q=(0.5,0.5,L)$ (signal) and $Q=(0.7,0.7,L)$ (background)
directions at $E=3,5.5$, and 7 meV, respectively. Inspection of
the data immediately suggests that the scattering is antiferromagnetically
correlated between the layers along the $c$-axis. To model the AF spin
correlations along the $c$-axis, we assume their structure factor is similar
to that of \cite{diallo10}: $f(Q_{z})=F^{2}(Q)[S_{0}+S_{1}\exp
(-(L-L_{0})^{2}/2\sigma ^{2})]$, where $F(Q)$ is the magnetic form factor 
\cite{ratcliff}, $L=Q_{z}c/2\pi $, $L_{0}=\pm 1,\pm 3,\cdots $, $\sigma $ is
the width of the Gaussian which gives the correlation length of the spin
excitations along the $c$-axis, $S_{0}$ and $S_{1}$ are fitting parameters
for constant magnetic rod scattering along any $L$ and maximum magnetic
intensity at odd values of $L$, respectively.

Figure 7 shows the temperature dependence of the magnetic scattering at $E =3$ meV along the $c$-axis. 
Starting from 2 K and 20 K in Fig.
7(a), we see two clear peaks centered around $Q=(0.5,0.5,1)$ and $(0.5,0.5,-1)$ above the background. 
Gaussian fits to these peaks give a $c$-axis spin correlation length of $\sim $14 \AA\ [Fig. 7(b)]. Upon increasing
temperature to 50 K and 70 K, the magnetic peaks at $Q=(0.5,0.5,1)$ and $(0.5,0.5,-1)$ reduce in intensity but the $c$-axis spin-spin correlations
appear not to be affected [Fig. 7(c)]. To test if the magnetic scattering
below $T_{N}$ simply follows the Bose population factor as expected for
simple spin waves excitations, we show in Fig. 7(d) the temperature
dependence of the magnetic scattering normalized to the scattering at 2 K.
The observed magnetic scattering clearly does not obey the Bose population
factor, suggesting that the spin excitations in the doped materials are not
simple spin waves, in contrast to the (undoped) parent compounds.

\begin{figure}[t]
\includegraphics[scale=.45]{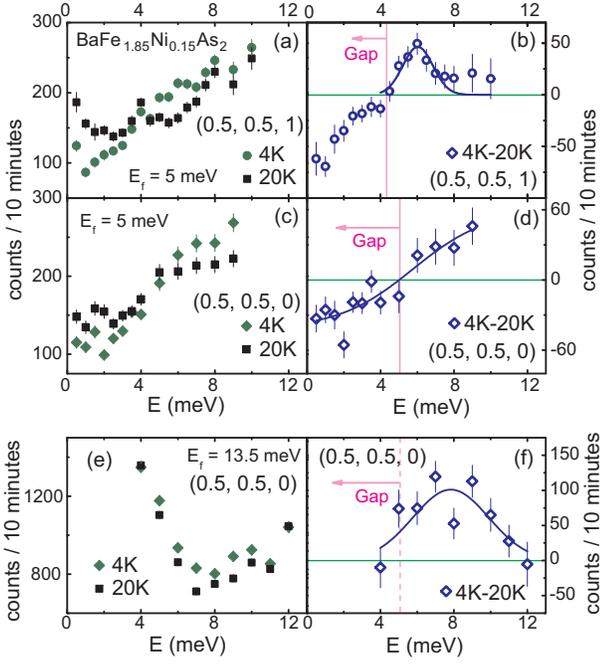}
\caption{(color online)
Summary of constant-$Q$ scans at various temperatures for BaFe$_{1.85}$Ni$_{0.15}$As$_2$ with $E_f=5$ meV
and 13.5 meV.
(a) Energy dependence of scattering at $Q=(0.5,0.5,1)$ above and below $T_c$.
The scattering shows a clear drop in intensity below $E=4$ meV 
and enhancement near $E=5$ meV.  These data suggest that
the effect of superconductivity is to open a low-energy spin
gap and form a neutron spin resonance at $E=5$ meV.  
(b) The temperature difference data confirm the formation of the
neutron spin resonance.
(c) Energy dependence of the scattering above and below $T_c$, now at 
$Q=(0.5,0.5,0)$.  One can only probe magnetic scattering below $E=9$ 
meV due to kinematic constraints with $E_F=5$ meV.  The scattering shows clear enhancement below
$T_c$ at energies around $E=8$ meV, as confirmed by the temperature difference plot
in (d).  (e)
Identical scan as that of (c) except we used $E_f=13.5$ meV.  Superconductivity induced
neutron spin resonance can now be seen around $E=8$ meV.
(f) The temperature difference plot confirms the presence of the resonance at 
$E=8$ meV.
 }
\end{figure}

To probe the $L$-dependence of the magnetic scattering at the neutron spin
resonance energies, we show in Figs. 8 and 9 constant-energy scans at $E =5.5$ meV and 7 meV, respectively. Consistent with constant-energy
scan data at $E =3$ meV (Fig. 7), the magnetic scattering is
still centered at $L=\pm 1,\pm 3$ positions and superconductivity has a
relatively small effect on the overall magnetic scattering. Comparison of
Figs. 2, 4, and 8, 9 reveals that the neutron spin resonance so clearly
illustrated in the temperature difference scattering is a rather subtle
effect that occurs at both $L=0$ and $L=1$ [Figs. 8(c) and 9(a)]. On warming
to 50 K and 70 K, the magnetic scattering decreases but the $c$-axis spin
correlation length of $\sim $14 \AA\ appears to be fairly temperature
independent. The solid lines in Fig. 8(a) and 9(b) show Gaussian fits to the
data which indicates that the decrease in the magnetic scattering along the $c$-axis direction falls off faster than just the Fe$^{2+}$ form factor \cite{ratcliff}. If we assume that the spins prefer to lie in the $a$-$b$ plane as is the case for
the AF undoped system, an additional reduction in intensity is expected due
to the neutron spin-Fe-spin orientation factor and this brings the curve in
reasonable agreement with the data.

\begin{figure}[t]
\includegraphics[scale=.45]{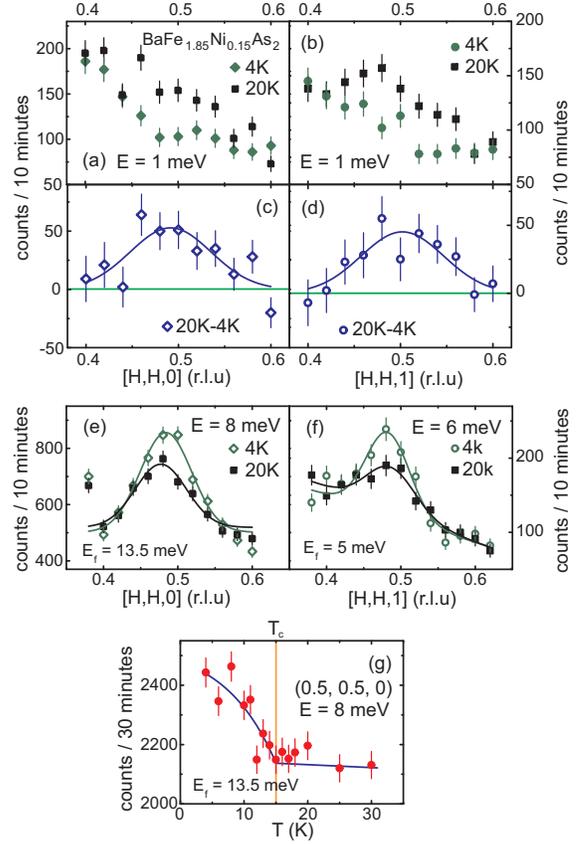}
\caption{(color online)
Summary of constant-energy scans at an energy below the spin gap energy and at the resonance energy
below and above $T_c$.
(a) $Q$-scan along the $[H,H,0]$ direction below and above $T_c$ at $E=1$ meV.
There is a clear peak centered at $(0.5,0.5,0)$ that disappears below $T_c$, thus suggesting
the opening of a clean spin-gap at $E=1$ meV and $(0.5,0.5,0)$.
(b) Similar scan along the $[H,H,1]$ direction, again indicating the opening a spin gap at
$E=1$ meV and $(0.5,0.5,1)$.
(c,d) Temperature difference plots confirm that superconductivity-induced spin gaps occur at
$(0.5,0.5,0)$ and $(0.5,0.5,1)$ positions.
(e) Constant-energy scans along the $[H,H,0]$ direction below and above $T_c$ at 
the resonance energy of $E=8$ meV.  (f) Similar scans at $E=6$ meV
along the $[H,H,1]$ direction.  In both cases, we find that superconductivity-induced
changes happen at the expected wave-vectors.
(g) Temperature dependence of the scattering at $E=8$ meV and $Q=(0.5,0.5,0)$.
The data show clear order-parameter-like increase below $T_c$, a hall mark of the neutron spin resonance.
 }
\end{figure}

Having described our comprehensive measurements on the underdoped BaFe$_{1.925}$Ni$_{0.075}$As$_{2}$, we now discuss inelastic neutron scattering
experiments on the overdoped BaFe$_{1.85}$Ni$_{0.15}$As$_{2}$ [Figs. 1(a),
1(c)], where the static AF order is completely suppressed. These
measurements were carried out on the PANDA cold triple-axis spectrometer.
Figure 10 summarizes the constant-$Q$ scans at $Q=(0.5,0.5,1)$ and $(0.5,0.5,0)$ below and above $T_{c}$. 
Using $E_{f}=5$ meV, we find in Figs. 10(a) and 10(b) that the neutron spin resonance occurs at $E=6$
meV for $Q=(0.5,0.5,1)$. Similar scans at $Q=(0.5,0.5,0)$ reveal clear
scattering intensity enhancement above $E =5$ meV [Figs. 10(c)
and 10(d)]. However, kinematic constraints with the $E_{f}=5$ meV
spectrometer configuration did not allow a conclusive determination of the
resonance energy. Figure 10(e) shows identical scans carried out with $E_{f}=13.5$ meV. Inspection of Figs. 10(e) and 10(f) indicates that the
resonance energy at $Q=(0.5,0.5,0)$ is now shifted to $E =8$ meV.
If we assume that the negative scattering in the temperature difference
spectra of Figs. 10(b) and 10(d) gives the onset of the spin gap (assuming background scattering is temperature independent between 2 K and 20 K), these
results suggest that the spin gap at $Q=(0.5,0.5,0)$ is larger than that at $Q=(0.5,0.5,1)$, consistent with the previous conclusions \cite{slli}. 

\begin{figure}[t]
\includegraphics[scale=.4]{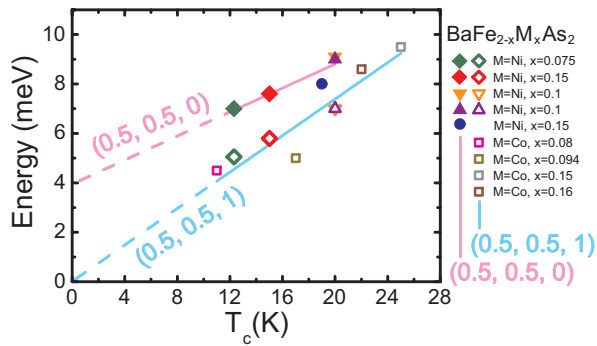}
\caption{(color online)
Summary of electron-doping dependence of the neutron spin resonance energies at $Q=(0.5,0.5,0)$ and $(0.5,0.5,1)$ as a function of $T_c$. The data for BaFe$_{2-x}$Ni$_{x}$As$_2$ were from Refs. \cite{chi,slli,harriger,jzhao09} and
present work.  The data for BaFe$_{2-x}$Ni$_{x}$As$_2$ were from Refs. \cite{lumsden,pratt,christianson09,inosov}.
The solid lines are linear fits to the data. 
 }
\end{figure}

To determine if the low-energy spin excitations have a clean spin gap like
those of the optimally doped BaFe$_{1.9}$Ni$_{0.1}$As$_{2}$ \cite{chi,slli},
we took constant-energy scans along the $[H,H,0]$ and $[H,H,1]$ directions
above and below $T_{c}$ for $E =1$ meV. Figures 11(a) and 11(b)
show that spin excitations of BaFe$_{1.85}$Ni$_{0.15}$As$_{2}$ are gapless
in the normal state for both $L=0$ and 1 rlu, but open a clean gap below $T_{c}$ at $E =1$ meV . These results are similar to the optimally
Ni-doped material, but are clearly different from underdoped BaFe$_{1.925}$Ni$_{0.075}$As$_{2}$ where there are no clean spin gap at $E =2.5$
meV [Fig. 5(a)]. Figures 11(e) and 11(f) show wave-vector scans at the
expected resonance energies for $Q=(0.5,0.5,0)$ and $(0.5,0.5,1)$. In both
cases, we find clear intensity enhancement below $T_{c}$. The temperature
dependent scattering at $Q=(0.5,0.5,0)$ and $E =8$ meV in Fig.
11(g) shows clear order-parameter-like intensity increase below $T_{c}$,
thus confirming the neutron spin resonance \cite{christianson,lumsden,chi,slli,pratt,christianson09,inosov}.

Finally, we summarize in Figure 12 the electron-doping dependence of the
neutron spin resonance at $Q=(0.5,0.5,0)$ and $(0.5,0.5,1)$ as a function of 
$T_{c}$ for both BaFe$_{2-x}$Ni$_{x}$As$_{2}$ \cite{chi,slli,harriger,jzhao09} and BaFe$_{2-x}$Co$_{x}$As$_{2}$ \cite{lumsden,pratt,christianson09,inosov}. For copper oxide high-$T_{c}$
superconductors, one of the hallmarks of the resonance is that its energy is
proportional to $T_{c}$ over a very wide temperature range \cite{wilson,jzhao07}. Since BaFe$_{2-x}$Ni$_{x}$As$_{2}$ has two resonances at
distinctively different energies, its dispersion along the $c$-axis is
related to the superconducting gap $\Delta _{0}$ and its deviation $\delta $
via $E (Q_{z})\sim 2\Delta _{0}-2\delta \left\vert \sin
(Q_{z}/2)\right\vert $ \cite{chi}. The observation of a linear relationship
for both mode energies and $T_{c}$ suggests that $\delta /\Delta
_{0}=[\omega (0.5,0.5,0)-\omega (0.5,0.5,1)]/\omega (0.5,0.5,0)$ is
approximately 0.28 and weakly Ni-doping dependent. Therefore, the ratio of
interplane ($J_{\perp }$) and intraplane ($J_{\parallel }$) AF coupling, $J_{\perp }/J_{\parallel }$, is weakly electron-doping dependent assuming
that the values of $\Delta _{0}$ and $\delta $ are proportional to $J_{\parallel }$ and $J_{\perp }$, respectively.

We now discuss the physical interpretation of the above results. In the
theory of spin-fluctuation-mediated superconductivity \cite{mazin,chubkov,fwang,cvetkovic,moreo}, the electron pairing arises from sign-reversed 
$S$-wave interband scattering between hole pockets centered at the $\Gamma$
point and electron pockets at the $M$ points [inset in Fig. 1(a)] \cite{maier1,maier2,korshunov,seo09}. One of the consequences of such
electron-hole pocket excitations is to induce a resonance peak at the AF
ordering wave vector $Q=(0.5,0.5,0)$ in the spin excitations spectrum. In
the strictly two-dimensional model, the energy of the resonance is at (or slightly less than) the addition of hole ($\Delta _{0}^{h}$) and electron ($\Delta _{0}^{e}$) superconducting gap energies ($\Delta _{0}=\Delta
_{0}^{h}+\Delta _{0}^{e}$). Our previous finding of three-dimensionality of
the resonance in optimally doped BaFe$_{1.9}$Ni$_{0.1}$As$_{2}$ \cite{chi,slli} suggests that the superconducting gap energy $\Delta _{0}$ should
be three-dimensional as well and sensitive on the $Q$ values along the $c$-axis. The new results reported in the present paper on underdoped BaFe$_{1.925}$Ni$_{0.075}$As$_{2}$ and overdoped BaFe$_{1.85}$Ni$_{0.15}$As$_{2}$
confirm the earlier conclusion, and reveal that the three-dimensional nature
of the superconducting gap is prevalent throughout the superconducting
dome. If spin excitations are mediating the electron pairing for
superconductivity, these results would suggest that the AF exchange coupling along the $c$-axis ($J_{\perp }$) 
contributes significantly to the electron pairing. Although
the overall spin excitations as a function of increasing electron doping transform into
quasi two-dimensional spin excitations rather rapidly as demonstrated by the
disappearing anisotropic spin gaps at $Q=(0.5,0.5,0)$ and $(0.5,0.5,1)$ with
increasing Ni-doping \cite{harriger}, the superconductivity-induced resonance retains its
three-dimensional character even in the overdoped regime. This means that
the superconducting electronic gaps in the iron-arsenic based materials are three-dimensional and 
quite different from that of the copper oxide superconductors.

\section{IV. Conclusions}

In summary, we have determined the doping evolution of the low-energy spin
excitations in BaFe$_{2-x}$Ni$_{x}$As$_{2}$ for both underdoped and
overdoped superconductors. In underdoped BaFe$_{1.925}$Ni$_{0.075}$As$_{2}$
we find that the appearance of bulk superconductivity is associated with the
appearance of a weak three-dimensional neutron spin resonance. The spectral
weight gain of the resonance below $T_{c}$ is a rather small portion of the
overall normal state magnetic scattering, and is compensated by opening a
weak pseudo spin gap and reduction in static magnetic moment. Our Ni-doping
dependent investigation of the spin gap and neutron spin resonance reveals
that the three-dimensional nature of the mode found earlier for the
optimally doped sample is a universal property of Ni-doped superconductors.
These results in turn suggest that AF spin excitations between the layers
are also important for the superconductivity of these materials.

\section{V. Acknowledgements}

We thank Jiangping Hu and Tao Xiang for helpful discussions.  
The neutron scattering part of this work at UT/ORNL is 
supported by the U.S. NSF No. DMR-0756568, and by the U.S. DOE, Division of Scientific
User Facilities.  The single crystal growth effort at UT is supported by
 U.S. DOE BES No. DE-FG02-05ER46202.  The single crystal growth and neutron scattering 
work at IOP is supported by Chinese Academy of Sciences and 973 Program (2010CB833102).


\end{document}